\documentclass[12pt,a4paper]{article}
\usepackage{amsmath,amsfonts,amsthm}
\allowdisplaybreaks[3]
\numberwithin{equation}{section}
\newtheorem{theorem}{Theorem}[section]

\newtheorem{corollary}[theorem]{Corollary}

\usepackage[body={16cm,23cm}]{geometry}
\newcommand{\fs}{{\mathbf f}}
\newcommand{\es}{{\mathbf e}}
\newcommand{\hs}{{\mathbf h}}
\newcommand{\Lf}{{\mathbf L}}
\newcommand\Lplus[1]{\mathcal{L}^{(#1,+)}}

\newcommand\Tplus[1]{\mathcal{T}^{+}_{#1} }
\newcommand\A[1]{\mathcal{A}^{(#1)}}
\newcommand\Q[1]{\mathcal{Q}^{(#1)}}
\begin{document}
\title{
The universal R-matrix and factorization of the L-operators related to the
Baxter Q-operators
}
\author{
Sergey Khoroshkin\thanks{
Institute of Theoretical and Experimental Physics, 117218 Moscow, Russia;
Higher School of Economics, Myasnitskaya 20, 101000, Moscow, Russia
}
 \and
Zengo Tsuboi\thanks{Present address: Department of Mathematics and Statistics, 
The University of Melbourne, Royal Parade, Parkville, Victoria 3010, Australia}}
\date{}
\maketitle
\begin{abstract}
We consider the `universal monodromy operators'  for the Baxter Q-operators.
 They  are given as images of the universal R-matrix in oscillator representation.
We find related universal factorization formulas in
   $U_{q}(\hat{sl}(2))$ case.
\end{abstract}
Keywords: Baxter Q-operator, universal R-matrix, L-operator, 
integrable system
\\
PACS number:  02.20.Uw, 02.30.Ik
\\
Journal Reference: J.\ Phys.\ A: Math.\ Theor.\ 47 (2014) 192003 
(Fast Track Communications)
\\
doi:10.1088/1751-8113/47/19/192003


\section{Introduction}
The Baxter Q-operators were introduced by Baxter \cite{Bax72},
and are considered to be one of the most powerful tools in quantum
integrable systems.
The T-operators (transfer matrices) are expressed in terms of
Q-operators.  Thus  Q-operators are fundamental objects.

In particular, Bazhanov, Lukyanov and Zamolodchikov
defined  Q-operators as the partial trace of 
the $L$-operator, which is a bosonic realization of the 
universal R-matrix, 
over some q-oscillator representations of the Borel subalgebra of
 a quantum affine algebra \cite{BLZ97}.
This `q-oscillator construction' of the Q-operators
was developed and generalized for
 higher rank algebras, superalgebras and so forth
(\cite{BHK02,Boos09,BT08,BGKNR10,T12} and references therein).
It is known 
that the q-oscillator representations for the Q-operators
are given as  limits of some representations of the Borel subalgebra of quantum affine algebra.
Recently, a systematic account on this from the point of view of the
representation theory was given and used in \cite{HJ11}.

The T and Q-operators are given as the traces of
(product of) the L-operators, which are images of the universal R-matrix.
In particular, L-operators for (infinite dimensional) Verma modules
 of the quantum affine algebra (or the Yangian)
factorize with respect to `L-operators for the Q-operators'.
Examples for such factorization formulas were given by a number of authors
(\cite{Derkachov,BLMS10} and references therein).
In this paper, we reconsider this phenomenon in relation to  properties of
the universal R-matrix and obtain a universal factorization formula,
which is independent of the quantum space.
Throughout this paper, we use  presentations of the universal R-matrix
from \cite{KT92,TK92}. We remark that
a detailed explanation on the evaluation of the universal R-matrix
 related to Q-operators,
which might be omitted or missing in earlier papers,
can be found in a recent series of papers \cite{BGKNR10}.
\section{Quantum algebras}
\subsection{$U_{q}(\hat{sl}(2))$}
The quantum affine algebra $U_{q}(\hat{sl}(2))$ is a Hopf algebra
generated by the generators
 $e_{i},f_{i},h_{i},d$, where
$i \in \{0,1 \}$.
We introduce the $q$-commutator
$[X,Y]_{q}=XY-q YX$. In particular,
$[X,Y]_{1}=[X,Y]$.
For $i,j \in \{0,1\}$, the
 defining relations of the algebra $U_{q}(\hat{sl}(2))$ are
given by
\begin{align}
\begin{split}
& [h_{i},h_{j}]=0, \quad [h_{i}, e_{j} ] =a_{ij} e_{j}, \quad
[h_{i}, f_{j} ] =-a_{ij} f_{j},
\\
&[e_{i},f_{j}]=\delta_{ij} \frac{q^{h_{i}} -q^{-h_{i}} }{q-q^{-1}},
\\
&[e_{i},[e_{i},[e_{i},e_{j}]_{q^{2}}]]_{q^{-2}}=
[f_{i},[f_{i},[f_{i},f_{j}]_{q^{-2}}]]_{q^{2}}=0
\qquad i \ne j ,
\end{split}
\end{align}
where $(a_{ij})_{0 \le i,j\le 1}$ is the
Cartan matrix
\begin{align}\nonumber
(a_{ij})_{0 \le i,j\le 1}=
\begin{pmatrix}
2& -2 \\
-2 & 2
\end{pmatrix}
.
\end{align}
We use the following co-product
  $ \Delta : U_{q}(\hat{sl}(2)) \to U_{q}(\hat{sl}(2)) \otimes U_{q}(\hat{sl}(2))$:
\begin{align}
\Delta (e_{i})&=e_{i} \otimes 1 + q^{-h_{i}} \otimes e_{i}, \nonumber\\
\Delta (f_{i})&=f_{i} \otimes q^{h_{i}} + 1 \otimes f_{i},\label{copro-h} \\
\Delta (h_{i})&=h_{i} \otimes 1 + 1 \otimes h_{i}. \nonumber
\end{align}
We will also use an opposite co-product defined by
\begin{align}
\Delta'=\sigma\circ \Delta,\qquad \sigma\circ
(X\otimes Y)=
Y\otimes X,\qquad X,Y\in U_{q}(\hat{sl}(2)).
\end{align}
We always assume zero value of the central element $h_{0}+h_{1}$.
Anti-pode, co-unit and grading element $d$ are not used  in this
paper.

The Borel subalgebra ${\mathcal B}_{+}$
(resp. ${\mathcal B}_{-}$) is generated by
$e_{i}, h_{i} $ (resp. $f_{i},h_{i}$), where
$i \in \{0,1 \}$.
For complex numbers $c_{i} \in {\mathbb C}$ which obey the relation
$\sum_{i=0}^{1} c_{i} =0$, the  transformation
\begin{align}
h_{i} \mapsto h_{i} +c_{i}, \qquad  i = 0,1,
 \label{shiftauto}
\end{align}
determines the shift automorphism $\tau_{c_{1}}$ of $ {\mathcal
B}_{+} $ (or of ${\mathcal B}_{-} $). Here we omit the unit element
multiplied by the above complex numbers.

There exists a unique element \cite{Dr85,KT92}
${\mathcal R}$ in a completion of $ {\mathcal B}_{+} \otimes {\mathcal B}_{-} $
called the universal R-matrix which satisfies the following
relations
\begin{align}
\Delta'(a)\ {\mathcal R}&={\mathcal R}\ \Delta(a)
\qquad \text{for} \quad \forall\ a\in U_{q}(\hat{sl}(2))\,   ,\nonumber\\
(\Delta\otimes 1)\,
{\mathcal R}&={\mathcal R}_{13}\, {\mathcal R}_{23}\, ,\label{R-def}\\
(1\otimes \Delta)\, {\mathcal R}&={\mathcal R}_{13}\,
{\mathcal R}_{12}\,\nonumber
\end{align}
where
${\cal R}_{12}={\cal R}\otimes 1$, ${\cal R}_{23}=1\otimes {\cal R}$,
${\cal R}_{13}=(\sigma\otimes 1)\, {\cal R}_{23}$.
The Yang-Baxter equation
\begin{align}
{\mathcal R}_{12}{\mathcal R}_{13}{\cal R}_{23}=
{\mathcal R}_{23}{\mathcal R}_{13}{\mathcal R}_{12}\ ,\label{YBE}
\end{align}
is a corollary of these relations \eqref{R-def}.
The universal R-matrix can be written in the form
\begin{align}
{\mathcal R}=\overline{{\mathcal R}}\ q^{\mathcal K},
\qquad {\mathcal
  K}=\frac{1}{2}h_{1}\otimes h_{1}. \label{R-red}
\end{align}
Here $\overline{{\mathcal R}}$  is the reduced universal $R$-matrix, which
is a series in $e_j\otimes 1$ and $1 \otimes f_j$
and does not contain Cartan elements
(to denote this symbolically, we will use a notation $\overline{{\mathcal R}}=
\overline{{\mathcal R}}(\{e_0\otimes 1, e_1 \otimes 1,
1 \otimes f_0 , 1 \otimes f_1 \})$).
Thus the reduced universal R-matrix is unchanged under the shift automorphism  $\tau_{c_{1}}$ of
 $ {\mathcal B}_{+}$, see (\ref{shiftauto}), while
the prefactor $\mathcal{K}$ is shifted as
\begin{align}
{\mathcal K}  \mapsto
{\mathcal K}   +
\frac{c_{1}}{2} (1 \otimes h_{1}).
\label{unRshift}
\end{align}

\subsection{$U_{q}(sl(2))$}
The algebra $U_{q}(sl(2))$ is  generated by the elements $E, F, H$.
The defining relations  are
\begin{align}
&
[H,E]=2E, \qquad [H,F]=-2F,
\nonumber
\\ &
[E, F] = \frac{q^{H} - q^{-H} }{q-q^{-1}} .
\label{HEF-sl2}
\end{align}
The following elements are central in $U_{q}(sl(2))$:
\begin{align}
C^{(r)}=FE+\frac{q^{H+1} + q^{-H-1}}{(q-q^{-1})^{2}}, \qquad
C^{(l)}=EF+\frac{q^{H-1} + q^{-H+1}}{(q-q^{-1})^{2}}
\end{align}
Due to the relations \eqref{HEF-sl2}, $C^{(r)}=C^{(l)}$ holds.

There is an evaluation map $\mathsf{ev}_{x}$:
$U_{q}(\hat{sl}(2)) \mapsto U_{q}(sl(2))$:
\begin{align}
\begin{split}
& e_{0} \mapsto F,  \qquad
 f_{0} \mapsto E, \qquad
 h_{0} \mapsto -H,
\\
& e_{1} \mapsto x E,  \qquad
f_{1} \mapsto x^{-1} F, \qquad
h_{1} \mapsto H,
\end{split}
\label{eva}
\end{align}
where $x  \in {\mathbb C}^*$ is a parameter.
Let $\pi_{\mu}^{+}$  be the Verma module over $U_{q}(sl(2))$ with the highest weight $\mu$. In a basis $\{v_{n} | n \in {\mathbb Z}_{\ge 0} \}$, we have
\begin{align}
H v_{n} =(\mu -2n)v_{n}, \quad
E v_{n}=[n]_{q}[\mu -n+1]_{q} v_{n-1},
\quad
 F v_{n}=v_{n+1} .
 \label{hwvglmn}
\end{align}
For $ \mu \in {\mathbb Z}_{\ge 0}$,
the module $\pi_{\mu}^{+}$ contains an invariant subspace isomorphic to  $\pi_{-\mu-2}^{+}$, and the quotient
 space  $\pi_{\mu}^{+} /\pi_{-\mu-2}^{+} $
is isomorphic to the finite dimensional irreducible module $\pi_{\mu} $
with the highest weight $\mu$. In particular,
$\pi_{1}(E)=E_{12}$, $\pi_{1}(F)=E_{21}$ and $\pi_{1}(H)=E_{11}-E_{22}$ gives the fundamental representation of $U_{q}(sl(2))$, where
$E_{ij}$ is a $2 \times 2$ matrix unit whose
$(k,l)$-element is $\delta_{i,k}\delta_{j,l}$.

Then the compositions
 $\pi_{\mu }^{+}(x)=\pi_{\mu }^{+} \circ \mathsf{ev}_{x}$  and
$\pi_{\mu }(x)=\pi_{\mu } \circ \mathsf{ev}_{x}$
give evaluation representations of $U_{q}(\hat{sl}(2))$.
%
%
\subsection{q-oscillator algebras}
We introduce two kinds of oscillator algebras $\mathrm{Osc}_{i}$ ($i=1,2$).
 They are generated by  the elements $\hs_{i},\es_{i}, \fs_{i}$ which are subject to the relations:
\begin{align}
\begin{split}
& [\hs_{1}, \hs_{1}]=0,
\qquad [\hs_{1},\es_{1}]=2\es_{1},
\qquad [\hs_{1},\fs_{1}]=-2\fs_{1},
\\
& \fs_{1}\es_{1}=q\frac{1-q^{\hs_{1}}}{( q-q^{-1})^{2} } ,
\qquad
\es_{1}\fs_{1}=q\frac{1-q^{\hs_{1}-2 }}{( q-q^{-1})^{2} },
\end{split}
\label{osc1}
\end{align}
\begin{align}
\begin{split}
& [\hs_{2}, \hs_{2}]=0,
\qquad [\hs_{2},\es_{2}]=2\es_{2},
\qquad [\hs_{2},\fs_{2}]=-2\fs_{2},
\\
& \fs_{2}\es_{2}=q^{-1}\frac{1-q^{-\hs_{2}}}{( q-q^{-1})^{2} } ,
\qquad
\es_{2}\fs_{2}=q^{-1}\frac{1-q^{-\hs_{2}+2 }}{( q-q^{-1})^{2} },
\end{split}
\label{osc2}
\end{align}
Note that $\mathrm{Osc}_{1}$ and  $\mathrm{Osc}_{2}$ can be swapped  by the transformation
$q \mapsto q^{-1} $.
One can derive the following corollaries of  \eqref{osc1} and \eqref{osc2}:
\begin{align}
& [\es_{1},\fs_{1}]=\frac{q^{\hs_{1} }}{q-q^{-1}},
\qquad
[\es_{2},\fs_{2}]=-\frac{q^{-\hs_{2} }}{q-q^{-1}},
\label{comosc1}
\\
& [\es_{1},\fs_{1}]_{q^{-2}}= \frac{1}{q-q^{-1}},
\qquad
[\es_{2},\fs_{2}]_{q^{2}} = -\frac{1}{q-q^{-1}}.
\label{comosc2}
\end{align}
Relations \eqref{comosc1} are nothing but contractions
\footnote{Contractions of a quantum algebra and its relation to
a q-oscillator algebra was discussed in \cite{Chaichian:1989rq}.}
 of the relations
 \eqref{HEF-sl2}.
On the other hand,
 the relations \eqref{comosc2}
are conditions that central elements take constant values.
 The following  limits of the generators of $U_{q}(sl(2))$ for the Verma module $\pi^{+}_{\mu}$
\begin{align}
\begin{split}
& \lim_{q^{-\mu} \to 0} (H-\mu)=\hs_{1},
\qquad
\lim_{q^{-\mu} \to 0} q^{-H-\mu}=0,  \\
& \lim_{q^{-\mu} \to 0} F=\fs_{1},
\qquad
 \lim_{q^{-\mu} \to 0} Eq^{-\mu}=\es_{1},
\end{split}
\label{limit1}
\end{align}
\begin{align}
\begin{split}
& \lim_{q^{\mu} \to 0} (H-\mu)= \hs_{2},
\qquad
\lim_{q^{\mu} \to 0} q^{H+\mu}=0,  \\
& \lim_{q^{\mu} \to 0} F=\fs_{2},
\qquad
 \lim_{q^{\mu} \to 0} Eq^{\mu}=\es_{2},
\end{split}
  \label{limit2}
\end{align}
realize the q-oscillator algebras $\mathrm{Osc}_{1}$ and
$\mathrm{Osc}_{2}$, respectively. By this construction, oscilator
algebras $\mathrm{Osc}_{1}$ and  $\mathrm{Osc}_{2}$ are given
together with irreducible highest weight representations $W_1$ and
$W_2$ of zero highest weight. The vector space of $W_1$ is generated
by vectors $v_n^{(1)}=\fs_1^nv_0^{(1)}$, and the vector space of
$W_2$ is generated by vectors $v^{(2)}_n=\fs_2^nv^{(2)}_0$, so that
$$\es_1 v_n^{(1)}=
[n]_q\frac {q^{-n+1}}{q-q^{-1}}v_{n-1}^{(1)},
\qquad 
\es_2 v_n^{(2)}=-[n]_q\frac {q^{n-1}}{q-q^{-1}}v^{(2)}_{n-1}.$$
%
Here $[n]_q=\frac{q^n-q^{-n}}{q-q^{-1}}$. 
\section{L-operators}
In the following we denote $\lambda=q-q^{-1}$. We set
\begin{align}
\Lf (x)& =
\begin{pmatrix}
q^{\frac{H}{2}} -q^{-1} x q^{-\frac{H}{2}} & \lambda F q^{-\frac{H}{2} } \\
\lambda x E q^{\frac{H}{2} } & q^{- \frac{H}{2}} -q^{-1} x q^{\frac{H}{2}}
 \end{pmatrix} ,
 \label{Lgen}
\end{align}
 We have
\begin{equation}\label{qdet}
\mathrm{q\,det}\Lf (x)=\Lf _{11}(x)\Lf _{22}(q^{-2}x)-
q^{-1}\Lf _{21}(x)\Lf _{12}(q^{-2}x)
=1-q^{-2}\lambda^{2} C^{(l)}x+q^{-4}x^{2}.
\end{equation}
Then
$$\Lf (x)= \phi(x)(\mathsf{ev}_{x} \otimes \pi_{1}(1) ) \mathcal{R},$$
and $\phi(x)$ solves the relation (\ref{qdet}), so that
$\phi(x)\phi(q^{-2}x)
=1-q^{-2}\lambda^{2} C^{(l)}x+q^{-4}x^{2} $.
Explicitly, it implies
$\phi(x)=\exp(\sum_{k=1}^{\infty}\frac{-C_{k}}{1+q^{2k}} \frac{x^{k}}{k} )$,
with central elements $C_k\in U_{q}(\hat{sl}(2))$, so that
$\pi^{+}_{\mu} (C_{k})=q^{(\mu+1)k}+q^{-(\mu+1)k}$.

Define $\Lf^{(1)}(x)$ and $\Lf ^{(2)}(x)$ as the following limits of
 homomorphic images of $\Lf (x)$:
\begin{align}
\Lf^{(1)}(x)& = \lim_{q^{-\mu } \to 0} (\pi^{+}_{\mu} \otimes 1)
\Lf (xq^{-\mu})q^{-\frac{\mu \otimes \pi_{1}(1)(h_{1})}{2}}
=
\begin{pmatrix}
q^{\frac{\hs_{1}}{2}}  & \lambda \fs_{1} q^{-\frac{\hs_{1}}{2} } \\
\lambda x \es_{1} q^{\frac{\hs_{1}}{2} } & q^{- \frac{\hs_{1}}{2}} -q^{-1} x q^{\frac{\hs_{1}}{2}}
 \end{pmatrix} ,
 \label{LQ1}
\\
\Lf ^{(2)}(x)& = \lim_{q^{\mu } \to 0} (\pi^{+}_{\mu} \otimes 1)
\Lf (xq^{\mu})q^{-\frac{\mu \otimes \pi_{1}(1)(h_{1})}{2}}
=
\begin{pmatrix}
q^{\frac{\hs_{2}}{2}} -q^{-1} x q^{-\frac{\hs_{2}}{2}} & \lambda \fs_{2} q^{-\frac{\hs_{2}}{2} } \\
\lambda x \es_{2} q^{\frac{\hs_{2}}{2} } & q^{- \frac{\hs_{2}}{2}}
 \end{pmatrix},
 \label{LQ2}
\end{align}
 where the limits \eqref{limit1} and \eqref{limit2} are
applied\footnote{The factor $q^{-\frac{\mu \otimes
\pi_{1}(1)(h_{1})}{2}}$ came from \eqref{unRshift} for $c_{1}=-\mu$.}
correspondingly to \eqref{Lgen}. The latter $L$-operators  can be
also presented as homomorphic images of the universal $R$-matrix
under the homomorphisms $\rho_x^{(i)}: \mathcal{B}_{+} \to
\mathrm{Osc}_{i}$ defined by the relations
\begin{align}\label{rho}
& \rho^{(i)}_{x}(e_{0})=\fs_{i}, \quad \rho^{(i)}_{x}(e_{1})=x \es_{i}, \quad \rho^{(i)}_{x}(h_{0})=-\hs_{i}, \quad \rho^{(i)}_{x}(h_{1})=\hs_{i},
\quad i=1,2.
\end{align}
Namely we have the relations $ \Lf^{(i)}(x)= \phi^{(i)}(x)(
\rho^{(i)}_{x}\otimes \pi_{1}(1) ) \mathcal{R}, $ where
$$\phi^{(1)}(x)=\exp(\sum_{k=1}^{\infty}\frac{-q^{k}}{1+q^{2k}}
\frac{x^{k}}{k} ), \qquad
\phi^{(2)}(x)=\exp(\sum_{k=1}^{\infty}\frac{-q^{-k}}{1+q^{2k}}
\frac{x^{k}}{k} ).$$ The L-operators \eqref{LQ1} and \eqref{LQ2} are
given by some contraction\footnote{A preliminary form of such
contraction was previously discussed in \cite{Bp} and developed in
\cite{talks,T12}.} of the original L-operator \eqref{Lgen} with
respect to the Cartan part (diagonal elements). We will also use
notations
\footnote{We find that $\A{2}(x)$ contains an infinite product of q-exponentials, 
while  $\A{1}(x)$ contains only finite (two) for the non-Cartan part.}
$$ \Lplus{\mu}(x)= (\pi_\mu^+(x)  \otimes 1){\mathcal R}\qquad \text{ and  }\qquad
\A{i}(x)=(\rho_x^{(i)} \otimes 1){\mathcal R}.$$
Thus $\Lf^{(i)}(x)= \phi^{(i)}(x)( 1 \otimes \pi_{1}(1) ) \A{i}(x) $.
\section{Representations of $\mathcal{B}_{+}$ and the factorization formula}
In this section, we reconsider the q-oscillator representations of
$\mathcal{B}_{+}$ introduced in \cite{BLZ97} (see also a detailed
explanation in \cite{Boos09}), and give a factorization formula of
the L-operator.  We will use the co-product in \cite{TK92,KT92} rather than
the one in \cite{BLZ97}.
We will repeatedly use formulas in Appendix B.

 Denote by $W_i(x)$ the pullbacks of basic highest weight representations of
 the oscillator algebras  by means of the homomorphisms $\rho_x^{(i)}: \mathcal{B}_{+} \to \mathrm{Osc}_{i}$.
 These are representations of  $\mathcal{B}_{+}$ in the spaces $W_i$.
In tensor product  $ W_{1}(xq^{\mu}) \otimes W_{2}(xq^{-\mu})$ we consider the basis $v_{n,m}$
\begin{align} \label{4.1}
v_{n,m}=\exp_{q^{-2}}
\left((q-q^{-1}\right) \es_{1} \otimes \fs_{2})\ v^{(1)}_{n} \otimes v^{(2)}_{m}
=\sum_{k=0}^{n}
\begin{pmatrix}
n \\ k
\end{pmatrix}_{q^{-2}}
v^{(1)}_{n-k} \otimes v^{(2)}_{m+k},
\end{align}
where we introduce notations
\begin{align}
 \exp_{q}(x)& =1+\sum_{k=1}^{\infty} \frac{x^{k}}{(k)_{q} ! },
\qquad
\begin{pmatrix}
n \\ k
\end{pmatrix}_{q}
= \frac{(n)_{q}!}{(k)_{q}! (n-k)_{q}! },
\nonumber
\\
(k)_{q}! & =(1)_{q}(2)_{q} \cdots (k)_{q} ,
\qquad
(k)_{q}=\frac{1-q^{k}}{1-q}.
 \nonumber
\end{align}
The action of the generators of ${\mathcal B}_{+}$ on these vectors
 is as follows:
\begin{align}
\begin{split}
\Delta(e_{0})v_{n,m}&=v_{n+1,m},
\\
\Delta(e_{1})v_{n,m}&=x[n]_{q} [\mu-n+1]_{q}v_{n-1,m} -\frac{xq^{-\mu +2n+m-1} [m]_{q}}{q-q^{-1}}v_{n,m-1} ,
\\
\Delta(h_{0})v_{n,m}&=2(n+m)v_{n,m},
\quad
\Delta(h_{1})v_{n,m}=-2(n+m)v_{n,m}.
\end{split}
\end{align}
Here we see a filtration by $m$, with quotients isomorphic to evaluation
Verma modules $\pi_{\mu}^{+}(x)$ shifted by automorphisms $\tau_{-2m}$.
Set $E_{1}=  (q^{\mu} -q^{-\mu -\hs_{1}} ) \es_{1} $,
$F_{1}=\fs_{1}$,
$H_{1}=\hs_{1} +\mu $, which
 realize the Verma module $\pi_{\mu}^{+}$ on the
basis $\{ v^{(1)}_{n} | n \in {\mathbb Z}_{\ge 0} \}$. In particular, $E_1,F_1$ and $H_1$ satisfy
$U_q(sl(2))$ relations (\ref{HEF-sl2}). We also set
$E_{2}=  \es_{2} $,
$H_{2}=\hs_{2} -\mu $.

Then, the above relations (\ref{4.1}) on vectors  $v^{(1)}_{n}\otimes v^{(2)}_{m}$
can be rewritten as the following identities on operators in $W_1\otimes W_2$
\begin{align}
\begin{split}
{\mathbf O}^{-1} \Delta(e_{0}) {\mathbf O}
&=F_{1} \otimes 1 ,
\qquad
 {\mathbf O}^{-1} \Delta(e_{1}) {\mathbf O}
=x E_{1}\otimes  1 +  q^{-H_{1}} \otimes x E_{2},
\\
{\mathbf O}^{-1} \Delta(h_{0}) {\mathbf O}
&= - H_{1} \otimes 1 - 1 \otimes H_{2},
\qquad
{\mathbf O}^{-1} \Delta(h_{1}) {\mathbf O}
 =H_{1} \otimes 1 + 1 \otimes H_{2},
\end{split}
\end{align}
where we set ${\mathbf O}=\exp_{q^{-2}} (\lambda \es_{1} \otimes \fs_{2}) $.
This should be interpreted as
\begin{align}
{\mathbf O}^{-1}
\left(
(\rho^{(1)}_{xq^{\mu}} \otimes \rho^{(2)}_{xq^{-\mu}} )
\Delta(\mathsf{a})
\right)
{\mathbf O}
 = (\mathsf{ev}^{(1)}_{x} \otimes \mathsf{ev}^{(2)}_{x}) \Delta(\mathsf{a}),
 \label{change-co-pro}
\end{align}
for $ \mathsf{a} \in {\mathcal B}_{+}$, where we introduced the
evaluation maps of $\mathcal{B}_{+}$ into two algebras, which may be
regarded as certain quotients of
 $U_q(sl(2))$:
\begin{align}
&
\mathsf{ev}^{(1)}_{x}(e_{0})=F_{1}, \quad  \mathsf{ev}^{(1)}_{x}(e_{1})=x E_{1}, \quad \mathsf{ev}^{(1)}_{x}(h_{0})=-H_{1}, \quad
\mathsf{ev}^{(1)}_{x}(h_{1})=H_{1},
\\
& \mathsf{ev}^{(2)}_{x}(e_{0})=0, \quad \mathsf{ev}^{(2)}_{x}(e_{1})=x E_{2}, \quad \mathsf{ev}^{(2)}_{x}(h_{0})=-H_{2},
\quad \mathsf{ev}^{(2)}_{x}(h_{1})=H_{2}.
\end{align}
Let us evaluate the universal R-matrix
$(\Delta \otimes 1){\mathcal R}={\mathcal R}_{13} {\mathcal R}_{23}$ on
$ W_{1}(xq^{\mu}) \otimes W_{2}(xq^{-\mu}) \otimes 1$.
\begin{multline}
{\mathbf O}_{12}^{-1}
\left(
(\rho^{(1)}_{xq^{\mu}} \otimes \rho^{(2)}_{xq^{-\mu}} \otimes 1)
(\Delta \otimes 1){\mathcal R}
\right)
{\mathbf O}_{12}
=
{\mathbf O}_{12}^{-1}
\left(
(\rho^{(1)}_{xq^{\mu}} \otimes \rho^{(2)}_{xq^{-\mu}} \otimes 1)
{\mathcal R}_{13} {\mathcal R}_{23}
\right)
{\mathbf O}_{12}
\\
=
{\mathbf O}_{12}^{-1}
\A{1}_{13}(x q^{\mu})
\A{2}_{23}(x q^{-\mu})
{\mathbf O}_{12}.
\label{fac1}
\end{multline}
On the other hand, taking note on \eqref{change-co-pro}, we obtain
\begin{multline}
{\mathbf O}_{12}^{-1}
\left(
(\rho^{(1)}_{xq^{\mu}} \otimes \rho^{(2)}_{xq^{-\mu}} \otimes 1)
(\Delta \otimes 1){\mathcal R}
\right)
{\mathbf O}_{12}
= (\mathsf{ev}^{(1)}_{x} \otimes \mathsf{ev}^{(2)}_{x} \otimes 1)
(\Delta \otimes 1){\mathcal R}
\\
 = (\mathsf{ev}^{(1)}_{x} \otimes \mathsf{ev}^{(2)}_{x} \otimes 1) {\mathcal R}_{13} {\mathcal R}_{23}
=
\Lplus{\mu}_{13}(x)
\left( (1 \otimes \mathsf{ev}^{(2)}_{x} \otimes 1){\mathcal R}_{23} \right) ,
\label{fac2}
\end{multline}
Combining \eqref{fac1} and \eqref{fac2}, we obtain the factorization formula
\begin{align}
{\mathbf O}_{12}^{-1}
\A{1}_{13}(x q^{\mu})
\A{2}_{23}(x q^{-\mu})
{\mathbf O}_{12}
=
\Lplus{\mu}_{13}(x)
\left( (1 \otimes \mathsf{ev}^{(2)}_{x} \otimes 1){\mathcal R}_{23} \right) .
\label{fac3}
\end{align}
Note that the co-product of the universal R-matrix has the form
$(\Delta \otimes 1){\mathcal R}=
\overline{\mathcal R}(\{ \Delta(e_{0}) \otimes 1, \Delta(e_{1}) \otimes 1 ,
 1\otimes 1 \otimes f_{0}, 1\otimes 1 \otimes f_{1} \})
q^{ \frac{ \Delta( h_{1} )\otimes h_{1} }{2}}$.
Then we obtain
\begin{multline}
 (1 \otimes \mathsf{ev}^{(2)}_{x} \otimes 1){\mathcal R}_{23}
=
\overline{\mathcal R}(\{ 0,
1 \otimes  x \es_{2} \otimes 1,
1 \otimes 1 \otimes f_{0}, 1 \otimes 1 \otimes f_{1} \})
q^{\frac{ 1 \otimes  (\hs_{2}-\mu)\otimes h_{1} }{2}}
= \\
=
\exp_{q^{-2}}  \left( \lambda \otimes x \es_{2}  \otimes f_{1} \right)
q^{\frac{ 1 \otimes  (\hs_{2}-\mu)\otimes h_{1} }{2}}  ,
\label{univR-triv}
\end{multline}
where we used an explicit form of the universal R-matrix \cite{KT92,TK92}.
The universal R-matrix is an infinite product of q-exponentials.
However, only one of them is non-trivial in this case
since the image of the generator $e_{0}$ is $0$.
From \eqref{fac3}, we arrive at the universal factorization formula
in $\mathcal{B}_{+}$, which is independent on the quantum space
 (which is a representation of $\mathcal{B}_{-}$).
\begin{theorem}
\begin{align}
{\mathbf O}_{12}^{-1}
\A{1}_{13}(x q^{\mu})
\A{2}_{23}(x q^{-\mu})
{\mathbf O}_{12}
=
\Lplus{\mu}_{13}(x)
\exp_{q^{-2}}  \left( \lambda  \otimes x \es_{2}  \otimes f_{1} \right)
q^{\frac{ 1 \otimes  (\hs_{2}-\mu)\otimes h_{1} }{2}}  .
\label{fac4}
\end{align}
\end{theorem}
In the right hand side, the first factor says that we have an
evaluation Verma module with highest weight $\mu$, the second factor
says that we have a filtration indexed by the powers of $\es_{2}$
and the third factor says that in each quotient of the filtration
the evaluation Verma module is given with a shift automorphism
$\tau_{-\mu}$ of $\mathcal{B}_{+}$. A similar factorization will
also occur  for $\mathcal{B}_{-}$ (see Appendix A). Let us evaluate
the quantum space (the third space) of \eqref{fac4} to the
fundamental evaluation representation $\pi_{1}(1)$ of
$\mathcal{B}_{-}$ ($f_{0}=E_{12}, f_{1}=E_{21},
h_{1}=-h_{0}=E_{11}-E_{22}$). Then we obtain the following
factorization formula for the L-operators \eqref{Lgen} for
$\pi_{\mu}^{+}$, \eqref{LQ1} and \eqref{LQ2}.
\begin{corollary}
\begin{multline}
{\mathbf O}_{12}^{-1}
\Lf^{(1)}_{13}(xq^{\mu})
\Lf^{(2)}_{23}(xq^{-\mu})
{\mathbf O}_{12} =
\\
=
\begin{pmatrix}
q^{\frac{H_{1} }{2}} -q^{-1} x q^{-\frac{H_{1}}{2}} & \lambda F_{1} q^{-\frac{H_{1}}{2} } \\
\lambda x E_{1} q^{\frac{H_{1} }{2} } & q^{- \frac{H_{1}}{2}} -q^{-1} x q^{\frac{H_{1}}{2}}
 \end{pmatrix}
\begin{pmatrix}
1 & 0 \\
\lambda x  \es_{2}  & 1
 \end{pmatrix}
\begin{pmatrix}
 q^{\frac{\hs_{2}-\mu }{2}} & 0 \\
0 & q^{-\frac{\hs_{2}-\mu }{2}}
 \end{pmatrix}
.
 \label{fac-mat}
\end{multline}
\end{corollary}
Here we use the identity
$$\pi_\mu^+\phi(x)=\phi^{(1)}(q^\mu x)\phi^{(2)}(q^{-\mu} x).$$
A factorization formula
 for the rational case ($q=1$), which is
similar to \eqref{fac-mat}, was given in \cite{BLMS10}.
\section{Baxter Q-operators}
In this section, we define  T and Q-operators, and mention a
factorization formula for the T-operators. We will use a convention
of \cite{T12} (or \cite{BT08}). We define the universal boundary
transformation as
\begin{align}
{\mathcal D} =q^{\varphi h_{1} }, \qquad \varphi \in {\mathbb C}.
\end{align}
The commutativity
$({\mathcal D} \otimes {\mathcal D}) {\mathcal R} =
{\mathcal R}  ({\mathcal D} \otimes {\mathcal D})$ follows
 from  \eqref{copro-h}. It implies the YB relation for the modified $R$-matrix
$ {\mathcal R} ({\mathcal D} \otimes 1)$.

We define the universal T-operator for the Verma module as
\begin{align}
\Tplus{\mu}(x)=
 (\mathrm{Tr}_{\pi^{+}_{\mu}(x) }  \otimes 1)
\left[ {\mathcal R} ({\mathcal D} \otimes 1) \right]
 =
 (\mathrm{Tr}_{\pi^{+}(x)} \otimes 1)
\left[ \overline{\mathcal R} \
\overline{\mathcal D} \right] ,
 \label{Tdef}
\end{align}
where
\begin{align}
\overline{\mathcal D} =q^{{\mathcal K} } ({\mathcal D} \otimes 1)
=(1 \otimes z)^{h_{1} \otimes 1},
\qquad
z=q^{\frac{1}{2} h_{1} + \varphi } .
\end{align}
We also define the universal Q-operators as
\begin{align}
 \Q{i}(x)=
{\mathcal \chi }_{i}^{-1}
 (\mathrm{Tr}_{W_{i}(x)} \otimes 1)
\left[ {\mathcal R} ({\mathcal D} \otimes 1) \right]
 =
 {\mathcal \chi }_{i}^{-1}
 (\mathrm{Tr}_{W_{i}(x)}  \otimes 1)
\left[ \overline{\mathcal R} \
\overline{\mathcal D} \right]
\qquad i=1,2,
\label{Qdef}
\end{align}
where we introduce normalization operators
\begin{align}
{\mathcal \chi }_{i}=
 (\mathrm{Tr}_{W_{i}(x)}  \otimes 1)
\left[ \overline{\mathcal D} \right]
=\frac{1}{1-z^{-2}} .
\end{align}
Eq.\ \eqref{Qdef} also have an expression
$  \Q{i}(x)= {\mathcal \chi }_{i}^{-1}
 (\mathrm{Tr}_{W_{i} }  \otimes 1)
\left[  \A{i}(x) \, \rho^{(i)}_{x} ( {\mathcal D} ) \otimes 1 \right] $.
Let us apply \eqref{change-co-pro} for ${\mathcal D} $, multiply this
by \eqref{fac4} and take the trace over the first and the second
space. Then taking note on the definitions \eqref{Tdef} and \eqref{Qdef},
we arrive at the factorization formula on the T-operator.
\begin{theorem} (cf.\ \cite{BLZ97})
\begin{align}
\Tplus{\mu}(x)=\frac{z^{\mu}}{1-z^{-2}}
\,  \Q{1}(xq^{\mu})\Q{2}(xq^{-\mu})
\end{align}
\end{theorem}
Note that $\Tplus{\mu}(0)=z^{\mu} /(1-z^{-2} )$
is the character of the Verma module $\pi_{\mu}^{+} $.
Thus the T-operator is the Baxterization of the character by the Baxter Q-operators in the sense of \cite{T09,T12}.
\section*{Appendix A: various factorization formulas }
\subsection*{Representations of ${\mathcal B}_{-}$}
\label{repBm}
\addcontentsline{toc}{section}{Appendix A}
\def\theequation{A\arabic{equation}}
\setcounter{equation}{0}
Set $F_{2}= \fs_{2} (q^{\mu} -q^{-\mu +\hs_{2}} )$,
$E_{2}=\es_{2}$,
$H_{2}=\hs_{2} -\mu $.
Let us introduce evaluation maps from ${\mathcal B}_{-}$
\begin{align}
&
\mathsf{ev}^{(1)}_{x}(f_{0})=0, \quad
\mathsf{ev}^{(1)}_{x}(f_{1})=x^{-1} F_{1}, \quad
\mathsf{ev}^{(1)}_{x}(h_{0})=-H_{1}, \quad
\mathsf{ev}^{(1)}_{x}(h_{1})=H_{1},  \label{ev1Bm} \\
&
\mathsf{ev}^{(2)}_{x}(f_{0})=E_{2}, \quad
\mathsf{ev}^{(2)}_{x}(f_{1})=x^{-1} F_{2}, \quad
\mathsf{ev}^{(2)}_{x}(h_{0})=-H_{2},  \quad
\mathsf{ev}^{(2)}_{x}(h_{1})=H_{2}, \label{ev2Bm}
\\
%
& \rho^{(i)}_{x}(f_{0})=\es_{i}, \quad
\rho^{(i)}_{x}(f_{1})=x^{-1} \fs_{i}, \quad
\rho^{(i)}_{x}(h_{0})=-\hs_{i}, \quad
\rho^{(i)}_{x}(h_{1})=\hs_{i},
 \quad i=1,2. \label{evoscBm}
\end{align}
Then we find that \eqref{change-co-pro}
also\footnote{However, this does not mean that the maps \eqref{evoscBm}
can be extended to the whole algebra  $U_{q}(\hat{sl}(2))$. They are regarded as
maps from contracted algebras on  $U_{q}(\hat{sl}(2))$ (cf.\ \cite{T12}).}
 hold true
for any ${\mathsf a} \in \mathcal{B}_{-} $
under \eqref{ev1Bm}-\eqref{evoscBm}.
From a similar argument on $\mathcal{B}_{+}$ case,
we arrive at the universal factorization formula on $\mathcal{B}_{-}$.
\begin{multline}
{\mathbf O}_{23}^{-1}
\left(
( 1 \otimes 1 \otimes \rho^{(2)}_{xq^{-\mu}} )
{\mathcal R}_{13}
\right)
\left(
(1 \otimes \rho^{(1)}_{xq^{\mu}} \otimes 1)
{\mathcal R}_{12}
\right)
{\mathbf O}_{23}
= \\
=
\left( ( 1 \otimes 1 \otimes \mathsf{ev}^{(2)}_{x}  ) {\mathcal R}_{13} \right)
  \exp_{q^{-2}}  \left( \lambda  e_{1}  \otimes x^{-1}\fs_{1} \otimes 1\right)
q^{\frac{ h_{1} \otimes (\hs_{1}+\mu) \otimes 1 }{2}} .
\label{fac4-2}
\end{multline}

\subsection*{The opposite case $W_{2}(xq^{-\mu}) \otimes W_{1}(xq^{\mu})$}
\label{repop}
Set
\begin{align}
& \tilde{E}_{1}= (q^{-\mu} -q^{\mu -\hs_{1}} )\es_{1},  \qquad
\tilde{F}_{1}= \fs_{1},
\qquad
\tilde{H}_{1}=\hs_{1} -\mu ,
\\
& \tilde{E}_{2}=\es_{2},  \qquad
\tilde{F}_{2}= \fs_{2} (q^{-\mu} -q^{\mu +\hs_{2}} ),
\qquad
\tilde{H}_{2}=\hs_{2} +\mu .
\end{align}
These generators with tilde can be obtained from the ones without tilde by $\mu \to -\mu$.
Let us introduce extended evaluation maps
\begin{align}
&
\tilde{\mathsf{ev}}^{(1)}_{x}(e_{0})=\tilde{F}_{1}, \quad  \tilde{\mathsf{ev}}^{(1)}_{x}(e_{1})=0, \quad \tilde{\mathsf{ev}}^{(1)}_{x}(h_{0})=-\tilde{H}_{1}, \quad
\tilde{\mathsf{ev}}^{(1)}_{x}(h_{1})=\tilde{H}_{1}, \\
&
\tilde{\mathsf{ev}}^{(1)}_{x}(f_{0})=\tilde{E}_{1},
\quad  \tilde{\mathsf{ev}}^{(1)}_{x}(f_{1})=x^{-1} \tilde{F}_{1}, \\
%
%
&\tilde{\mathsf{ev}}^{(2)}_{x}(e_{0})=\tilde{F}_{2}, \quad \tilde{\mathsf{ev}}^{(2)}_{x}(e_{1})=x \tilde{E}_{2}, \quad
\tilde{\mathsf{ev}}^{(2)}_{x}(h_{0})=-\tilde{H}_{2},  \quad
\tilde{\mathsf{ev}}^{(2)}_{x}(h_{1})=\tilde{H}_{2},
\\
& \tilde{\mathsf{ev}}^{(2)}_{x}(f_{0})=\tilde{E}_{2}, \quad \tilde{\mathsf{ev}}^{(2)}_{x}(f_{1})=0.
\end{align}
Then we obtain
\begin{multline}
{\mathbf O}_{21}^{-1}
q^{-\frac{\mu(1 \otimes \hs_{1} -\hs_{2} \otimes 1 )}{2}}
\left(
(\rho^{(2)}_{xq^{-\mu}} \otimes \rho^{(1)}_{xq^{\mu}} )
\Delta(\mathsf{a})
\right)
q^{\frac{\mu(1 \otimes \hs_{1} -\hs_{2} \otimes 1 )}{2}}
{\mathbf O}_{21}
=
(\tilde{\mathsf{ev}}^{(2)}_{x} \otimes \tilde{\mathsf{ev}}^{(1)}_{x}) \Delta(\mathsf{a}),
\\
  \mathsf{a} \in {\mathcal B}_{+} \quad \mathrm{or} \quad
\mathsf{a} \in {\mathcal B}_{-} .
 \label{change-co-pro3}
\end{multline}
Let us evaluate the universal R-matrix
$(\Delta \otimes 1){\mathcal R}={\mathcal R}_{13} {\mathcal R}_{23}$ on
$ W_{2}(xq^{-\mu}) \otimes W_{1}(xq^{\mu}) \otimes 1$.
From \eqref{change-co-pro3}, we arrive at the universal factorization formula on $\mathcal{B}_{+}$
\begin{multline}
{\mathbf O}_{21}^{-1}
q^{-\frac{\mu(1 \otimes \hs_{1} -\hs_{2} \otimes 1 ) \otimes 1}{2}}
\left(
(\rho^{(2)}_{xq^{-\mu}} \otimes 1 \otimes 1)
{\mathcal R}_{13}
\right)\cdot
\left(
(1 \otimes \rho^{(1)}_{xq^{\mu}} \otimes 1)
{\mathcal R}_{23}
\right)
q^{\frac{\mu(1 \otimes \hs_{1} -\hs_{2} \otimes 1 ) \otimes 1 }{2}}
{\mathbf O}_{21}
= \\
=
\left( (\tilde{\mathsf{ev}}^{(2)}_{x} \otimes 1 \otimes 1) {\mathcal R}_{13} \right)
\exp_{q^{-2}}  \left( \lambda  \otimes \fs_{1}  \otimes f_{0} \right)
q^{\frac{ 1 \otimes  (\hs_{1}-\mu)\otimes h_{1} }{2}}  .
\label{fac4-3}
\end{multline}
\subsection*{Factorization on both $\mathcal{B}_{+}$ and $\mathcal{B}_{-}$}
\label{facBpm}
We expect the following factorization formula with respect to four kinds
of images of the universal $R$-matrix:
\begin{multline}
{\mathbf O}_{34}^{-1} {\mathbf O}_{12}^{-1}
\left(
(\rho^{(1)}_{xq^{\mu}} \otimes 1 \otimes 1 \otimes \rho^{(2)}_{yq^{-\nu }} )
{\mathcal R}_{14}
\right)
\left(
(\rho^{(1)}_{xq^{\mu}} \otimes 1 \otimes  \rho^{(1)}_{yq^{\nu }} \otimes 1)
{\mathcal R}_{13}
\right)
\times
\\
\times
\left(
(1 \otimes \rho^{(2)}_{xq^{-\mu}} \otimes 1 \otimes \rho^{(2)}_{yq^{-\nu }})
{\mathcal R}_{24}
\right)
\left(
(1 \otimes \rho^{(2)}_{xq^{-\mu}} \otimes \rho^{(1)}_{yq^{\nu }} \otimes 1)
{\mathcal R}_{23}
\right)
{\mathbf O}_{12} {\mathbf O}_{34}
= \\
=
\left( (\mathsf{ev}^{(1)}_{x} \otimes 1 \otimes 1 \otimes \mathsf{ev}^{(2)}_{y} ) {\mathcal R}_{14} \right)
  \exp_{q^{-2}}  \left( \lambda  x (q^{\mu } -q^{-\mu  - \hs_{1} } ) 
\es_{1}  \otimes 1 \otimes y^{-1}\fs_{1}
\otimes 1\right)
\\
\times 
\exp_{q^{-2}}  \left( \lambda  \otimes x \es_{2}  \otimes 
 1 \otimes y^{-1} \fs_{2} (q^{\nu } -q^{-\nu + \hs_{2}} ) \right)
\exp_{q^{-2}}  \left( \lambda  q^{-\hs_{1} -\mu }  \otimes x \es_{2}  \otimes
 y^{-1} \fs_{1} \otimes q^{\hs_{2} -\nu }   \right)
\\
\times
q^{\frac{(\hs_{1}+\mu ) \otimes 1 \otimes (\hs_{1}+\nu ) \otimes 1 }{2}
+
\frac{ 1 \otimes  (\hs_{2}-\mu)\otimes 
 ( (\hs_{1} +\nu) \otimes 1 +1 \otimes ( \hs_{2} -\nu) ) }{2}} ,
\label{facfour}
\end{multline}
where $\mathsf{ev}^{(2)}_{y} $ is defined by \eqref{ev2Bm} replacing $(\mu,x)$ with $(\nu,y)$. 

\section*{Appendix B: q-analogue of H'Adamar formula}
\label{qHA}
\addcontentsline{toc}{section}{Appendix B}
\def\theequation{B\arabic{equation}}
\setcounter{equation}{0}
Here we mention formulas derived from the H'Adamar identity
\cite{KT91}
\begin{align}
\exp_{q} ({\mathbf A}) {\mathbf B} \exp_{q}^{-1} ({\mathbf A})=
\sum_{k=0}^{\infty} \frac{{\mathbf B}_{k}}{(k)_{q} !},
\end{align}
where ${\mathbf B}_{0}={\mathbf B}$, ${\mathbf B}_{k+1}=[{\mathbf A},{\mathbf B}_{k}]_{q^{k}}$.
Set ${\mathbf O}=\exp_{q^{-2}} (\lambda \es_{1} \otimes \fs_{2}) $.
To simplify the notation, we will abbreviate
 ${\mathbf o}_{1} \otimes 1$ and  $1 \otimes {\mathbf o}_{2} $ as
 ${\mathbf o}_{1}$ and ${\mathbf o}_{2}$ respectively
for any element ${\mathbf o}$ of the oscillator algebra.
Then we obtain (as formal series on $\es_{1} \fs_{2} $)
\begin{align}
\begin{split}
& {\mathbf O}^{-1} \es_{1} {\mathbf O}=\es_{1},
\qquad
{\mathbf O}^{-1} \es_{2} {\mathbf O}=\es_{2}-(1-q \lambda^{2} \es_{1} \fs_{2})^{-1} \es_{1}q^{-\hs_{2}},
\\
& {\mathbf O}^{-1} \fs_{1} {\mathbf O}=\fs_{1}-(1-q \lambda^{2} \es_{1} \fs_{2})^{-1} \fs_{2}q^{\hs_{1}} ,
\qquad
  {\mathbf O}^{-1} \fs_{2} {\mathbf O}=\fs_{2},
\\
& {\mathbf O}^{-1} q^{ \alpha  \hs_{1} } {\mathbf O}=
\frac{(q^{2\alpha +1} \lambda^2 \es_{1} \fs_{2}; q^{2} )_{\infty} }{ (q \lambda^2 \es_{1} \fs_{2}; q^{2} )_{\infty} } q^{ \alpha \hs_{1} },
\qquad
{\mathbf O}^{-1} q^{ \alpha  \hs_{2} } {\mathbf O}=
\frac{(q^{-2\alpha +1} \lambda^2 \es_{1} \fs_{2}; q^{2} )_{\infty} }{ (q \lambda^2 \es_{1} \fs_{2}; q^{2} )_{\infty} } q^{ \alpha \hs_{2} },
\end{split}
\label{Hada}
\end{align}
where $\alpha \in \mathbb{C}$, $(a;q)_{\infty}=\prod_{k=0}^{\infty}(1-aq^{k})$.
From the last two identities, one can derive the following relations.
\begin{align}
& {\mathbf O}^{-1} q^{ \alpha ( \hs_{1} +\hs_{2} )} {\mathbf O}= q^{ \alpha ( \hs_{1} +\hs_{2} )},
\qquad
{\mathbf O}^{-1} q^{  \pm \frac{\hs_{1}- \hs_{2}}{2} } {\mathbf O}=
 (1-q^{\pm 1} \lambda^2 \es_{1} \fs_{2})^{\mp 1}
q^{  \pm \frac{\hs_{1}- \hs_{2}}{2} } .
\end{align}
One can calculate the inverse of \eqref{Hada}.
\begin{align}
\begin{split}
& {\mathbf O}\es_{1} {\mathbf O}^{-1} =\es_{1},
\qquad
{\mathbf O}\es_{2} {\mathbf O}^{-1} =\es_{2}+ \es_{1}q^{-\hs_{2}},
\\
& {\mathbf O} \fs_{1} {\mathbf O}^{-1} =\fs_{1}+\fs_{2}q^{\hs_{1}} ,
\qquad
  {\mathbf O} \fs_{2} {\mathbf O}^{-1} =\fs_{2},
\\
& {\mathbf O} q^{ \alpha  \hs_{1} } {\mathbf O}^{-1} =
\frac{(q^{2\alpha -1} \lambda^2 \es_{1} \fs_{2}; q^{-2} )_{\infty} }{ (q^{-1} \lambda^{2}\es_{1} \fs_{2}; q^{-2} )_{\infty} } q^{ \alpha \hs_{1} },
\qquad
{\mathbf O} q^{ \alpha  \hs_{2} } {\mathbf O}^{-1} =
\frac{(q^{-2\alpha -1} \lambda^2 \es_{1} \fs_{2}; q^{-2} )_{\infty} }{ (q^{-1} \lambda^2 \es_{1} \fs_{2}; q^{-2} )_{\infty} } q^{ \alpha \hs_{2} }.
\end{split}
\end{align}
\section*{Acknowledgments}
The authors would like to thank Vladimir Bazhanov for useful
discussions. The work of S.K. was supported by RFBR grant
14-01-00547. The work of Z.T. was supported by the Australian
Research Council. A main part of this paper was written when he was
at the department of theoretical physics, research school of physics
and engineering, Australian national university.

\end{document}